\documentclass[prb,preprint,a4paper,showpacs]{revtex4} 
\usepackage{amsmath}
\usepackage{epsfig}
\begin{document}
\title{\bf Large diamagnetic persistent currents}
\author{Sheelan Sengupta Chowdhury$^1$, P. Singha Deo$^1$, Ashim Kumar Roy$^2$ and M. Manninen$^3$}
\affiliation{$^1$Unit for Nanoscience and Technology, S. N. Bose National Centre for Basic Sciences, Sector-III, Block-JD, Salt Lake, Kolkata-700098, India\\
$^2$Physics and applied mathematics unit, Indian Statistical Institute, 203, B. T. Road, Kolkata-700108, India\\
$^3$Nanoscience center, Department of Physics, University of Jyvaskyla, P.O. Box-35, 40014, Jyvaskyla, Finland.}
\begin{abstract}
In multichannel rings, evanescent modes will always co-exist with propagating
modes. The evanescent modes can carry a very large diamagnetic persistent
current that can oscillate with energy and are very sensitive to impurity
scattering. This provides a natural explanation for the large diamagnetic
persistent currents observed in experiments.
\end{abstract}
\maketitle
\begin{figure}
\centering{\psfig{figure=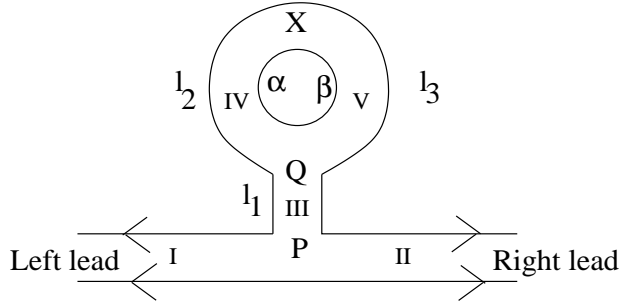,width=8cm,height=4cm,angle=0}}
\caption{A ring connected with an infinite wire. A $\delta$-potential impurity
is at position $X$.}\label{figsys}
\end{figure}

B\"{u}ttiker, Imry and Landauer first suggested the possibility of observing
persistent current in normal metal or semiconducting rings threaded by an
Aharonov-Bohm flux \cite{mb83}. This current is an equilibrium property of
the ring, given by the flux derivative of the total energy of the ring. Since
then several experiments have been done that confirm the existence of
persistent current in such rings
\cite{vc91,dm93,emqj01,wr01,lpl90,br95,rd02,rdrb02} through magnetization
measurements. However, the nature of these currents are quite different from
what is expected theoretically \cite{gm90}$^,$\cite{va90}. While earlier
experiments \cite{vc91,dm93,emqj01,wr01,lpl90,br95,rd02} had some ambiguity,
recent experiments made on an ensemble of $10^5$ rings have made very careful
measurements of the sign (positive implies diamagnetic and negative implies
paramagnetic) and periodicity of the persistent current \cite{rdrb02}. If an
ensemble of rings is taken, one can calculate an ensemble average over the
number of electrons in different rings or over disorder, or over both
\cite{gm90}. In such cases one finds that the ensemble average has $\phi_0/2$
( or $hc/2e$) periodicity and the low field persistent current is paramagnetic
in nature. One can also take a fixed chemical potential and average over
disorder. Here again one can calculate to find paramagnetic persistent current
with $\phi_0/2$ periodicity \cite{va90}. Whereas, the experiment \cite{rdrb02}
shows a persistent current, that has $\phi_0/2$ periodicity but diamagnetic in
nature at low fields and also of a large magnitude ($10$ to $100$ times larger
than that theoretically estimated in the above mentioned models of ensemble
averaging).
\vskip .2in
Any quantity that is very sensitive to disorder will average to zero. But the
second harmonic do not obey this rule and gives nonzero value. This is
essentially because the second harmonic consists of time reversed trajectories
and disorder configuration does not change the observed quantity randomly. This
is very robust and manifests in a variety of phenomena briefly described below.
Weak localization in disordered metallic or semi-conducting samples occur
because of this. Forward scattering probability beyond a certain length turns
out to be negligibly small while the back scattering arising due to time
reversed trajectories always interfere constructively, irrespective of disorder
configuration. As a consequence the Aronov-Altshuler-Spivak weak localization
correction to conductance has $\phi_0/2$ periodicity \cite{bla81}. Also the
response of a long cylinder to an applied magnetic field turns out to have
$\phi_0/2$ periodicity \cite{dys81}. $\phi_0/2$ periodicity of ensemble
averaged persistent current is due to the same reason that the first harmonic
averages to zero while the second does not. 
\vskip .2in
The first attempt to explain the discrepancy in sign and magnitude is based on
repulsive interactions between electrons \cite{va90}$^,$\cite{as91}. This did
not turn out to be the correct mechanism because this yields a paramagnetic
response at low fields whereas recent experiment has conclusively shown that
the observed response is diamagnetic. For a recent analysis of the effect of
disorder and interactions, we refer [15]. A more recent attempt to explain the
experimental discrepancy is based on additional currents that may be generated
in rings due to the rectification of a high frequency non-equilibrium noise
~\cite{vek00}. This mechanism can give a diamagnetic current in absence of spin
orbit coupling and a paramagnetic response in presence of spin orbit coupling.
Recent experiment \cite{rdrb02} has also ruled out this explanation as
paramagnetic response could not be observed in presence of strong spin orbit
coupling. The origin of a high frequency non-equilibrium noise also seems to be
unclear.
\vskip .2in
All experiments have been done at finite temperatures. There are thermal
effects wherein an electron can get energy from some collision and get excited
to higher states. Such inelastic processes will not destroy the persistent
currents. Persistent currents are actually observed in networks, where total
length is much greater than the inelastic mean free path \cite{mp97}. When a
mechanism for excitation is present then evanescent modes can be excited.
It has been shown that in a one dimensional (1D) ring, evanescent modes can
carry a diamagnetic persistent current. It has a very small magnitude compared
to persistent current in propagating modes and cannot exhibit $\phi_0/2$
periodicity as it is not sensitive to disorder. Rings used in the experiments
have a finite thickness and are referred to as quasi-one dimensional (Q1D)
rings. In this work we show that in Q1D, evanescent modes can carry very large
diamagnetic persistent current that are comparable to that of propagating modes
and are as sensitive to disorder as that due to the propagating modes. So this
mechanism is a natural explanation for the observed diamagnetic persistent
current.
\vskip .2in
In this work we use a simple technique to excite evanescent modes. We consider
the ring to be coupled to an infinite wire as schematically shown in Figure
\ref{figsys}. This basically constitutes an open system and it is known that it
can simulate the effects of inelastic collisions and thermal effects
\cite{mb85}. We shall see in our mathematical analysis how evanescent modes are
excited in this system very naturally. We consider two modes of propagation as
the results can be generalized to any number of modes. There is a 
$\delta$-potential impurity present in the ring at any arbitrary position $X$
[Fig \ref{figsys}]. We apply Aharonov-Bohm flux $\phi$ through the ring,
perpendicular to the plane of the paper. The Schr\"{o}dinger equation for a Q1D
wire in presence of a $\delta$-potential at $x=0$, $y=y_i$ is
\begin{eqnarray}
[-\frac{\hbar^2}{2 m}(\frac{\partial^2}{\partial x^2}+\frac{\partial^2}{\partial y^2})+V_c(y)+\gamma \delta(y-y_i)\delta(x)]\Psi(x,y) &=& E\Psi(x,y)\nonumber\\
\label{eqse}
\end{eqnarray}
Here $V_c$ is the confinement potential making up the quantum wires in Figure
\ref{figsys}. Solutions to Schr\"{o}dinger equation is a ring geometry can be
obtained by applying periodic boundary conditions to Eqn. \ref{eqse}. The
magnetic field just appears as a phase of $\Psi (x,y)$ that will be accounted
for while applying boundary conditions. Away from the scattering regions
Eqn. \ref{eqse} can be separated as
\begin{eqnarray}
-\frac{\hbar^2}{2 m} \frac{d^2\psi(x)}{dx^2} &=& \frac{\hbar^2k^2}{2m}\psi(x)
\label{eqsep1}
\end{eqnarray}
\noindent
and
\begin{eqnarray}
[-\frac{\hbar^2}{2 m} \frac{d^2}{dy^2}+V_c(y)] \chi_n(y) &=& E_n\chi_n(y)
\label{eqsep2}
\end{eqnarray}
Here we take $V_c$ to be a square well potential of width $W$ that gives
$\chi_n(y)=\sin [\frac{n\pi}{W}(y+\frac{W}{2})]$. In the first mode,
$k_1=\sqrt{\frac{2mE}{\hbar^2}-\frac{\pi^2}{W^2}}$ and in the second mode
$k_2=\sqrt{\frac{2mE}{\hbar^2}-\frac{4 \pi^2}{W^2}}$ are the propagating
wave-vectors. $m$ is the electron mass, $E$ is the electron energy
and $W$ is the width of the quantum wire. When electrons are incident along
region $I$ (in Fig \ref{figsys}) in the first mode the scattering problem can
be solved exactly. The solution to Eqn. \ref{eqsep1} in region $I$ becomes
\begin{eqnarray}
\psi_I & =& \frac{1}{\sqrt{k_1}}e^{ik_1x}+\frac{r^\prime_{11}}{\sqrt{k_1}} e^{-ik_1x}+\frac{r^\prime_{12}}{\sqrt{k_2}} e^{-ik_2x}
\end{eqnarray}
Similarly, in region $II$, $III$, $IV$ and $V$ we get
\begin{eqnarray}
\psi_{II} & = & \frac{g^\prime_{11}}{\sqrt{k_1}} e^{ik_1x}+ \frac{g^\prime_{12}}{\sqrt{k_2}} e^{ik_2x}\\
\psi_{III} & = & \frac{A e^{ik_1y}}{\sqrt{k_1}} + \frac{B e^{-ik_1y}}{\sqrt{k_1}}+\frac{C e^{ik_2y}}{\sqrt{k_2}}\nonumber\\
& & +\frac{D e^{-ik_2y}}{\sqrt{k_2}}\\
\psi_{IV} & = & \frac{E e^{ik_1z}}{\sqrt{k_1}} + \frac{F e^{-ik_1z}}{\sqrt{k_1}}+ \frac{G e^{ik_2z}}{\sqrt{k_2}}\nonumber\\
& & + \frac{H e^{-ik_2z}}{\sqrt{k_2}}\\
\psi_V & = & \frac{J e^{ik_1(z-l_2)}}{\sqrt{k_1}} +\frac{K e^{-ik_1(z-l_2)}}{\sqrt{k_1}}+\frac{L e^{ik_2(z-l_2)}}{\sqrt{k_2}}\nonumber\\
& & +\frac{M e^{-ik_2(z-l_2)}}{\sqrt{k_2}}
\end{eqnarray}
\noindent
The lead is along $x$ direction and region $III$ is along $y$ direction. $z$ is
used to denote coordinate inside the ring, the geometry of the ring being taken
care of by applying periodic boundary conditions to $z$. $r^\prime_{11}$,
$r^\prime_{12}$, $g^\prime_{11}$ and $g^\prime_{12}$ are the scattering matrix
elements and $A$, $B$, $C$, $D$, $E$, $F$, $G$, $H$, $J$, $K$, $L$, $M$ are to
be determined by mode matching.
\vskip .2in
Note that at $P$ and $Q$ we have a three legged junction that is schematically
shown in Fig. $2$. So far a popularly used scattering matrix for a three-legged
junction is \cite{mbyi83}
\begin{eqnarray}
S_U &=& \left(\begin{array}{cccccc}
-(a_s+b_s) & 0 & \sqrt{\epsilon} & 0 & \sqrt{\epsilon} & 0\\
0 & -(a_s+b_s) & 0 & \sqrt{\epsilon} & 0 & \sqrt{\epsilon}\\
\sqrt{\epsilon} & 0 & a_s & 0 & b_s & 0\\
0 & \sqrt{\epsilon} & 0 & a_s & 0 & b_s\\
\sqrt{\epsilon} & 0 & b_s & 0 & a_s & 0\\
0 & \sqrt{\epsilon} & 0 & b_s & 0 & a_s
\end{array}\right)\nonumber\\
\label{eqsmu}
\end{eqnarray}
with $a_s=\frac{1}{2}(\sqrt{1-2\epsilon}-1)$,
$b_s=\frac{1}{2}(\sqrt{1-2\epsilon}+1)$ and $0 < \epsilon < 0.5$. Such
junction $S$-matrix does not include channel mixing and can not account for
any contribution from evanescent modes. 
\vskip .2in
In this work we propose a three-legged junction scattering matrix $S_J$ for a
two channel quantum wire that can be easily generalized to any number of
channels. It is given by
\begin{eqnarray}
S_J &=& \left (\begin{array}{cccccc}
r_{11} & r_{12} & g_{11} & g_{12} & f_{11} & f_{12}\\
r_{21} &  r_{22} & g_{21} & g_{22} & f_{21} & f_{22}\\
g_{11} & g_{12} & r_{11} & r_{12} & f_{11} & f_{12}\\
g_{21} &  g_{22} & r_{21} & r_{22} & f_{21} & f_{22}\\
f_{11} & f_{12} & f_{11} & f_{12} & r_{11} & r_{12}\\
f_{21} &  f_{22} & f_{21} & f_{22} & r_{21} & r_{22}
\end{array} \right)
\label{eqsmj}
\end{eqnarray}
where
\begin{eqnarray}
r_{11} &=& -\frac{3k_2+k_1}{3 k_1+3 k_2}\nonumber\\
g_{11} = f_{11} &=& \frac{2 k_1}{3 k_1+3 k_2}\nonumber\\
r_{12} = g_{12} = f_{12} &=& \sqrt{\frac{k_2}{k_1}}\frac{2 k_1}{3 k_1+3 k_2}\nonumber\\
r_{22} &=& -\frac{3k_1+k_2}{3 k_1+3 k_2}\nonumber\\
g_{22}=f_{22} &=& \frac{2 k_2}{3 k_1+3 k_2}\nonumber\\
r_{21} = g_{21} = f_{21} &=& \sqrt{\frac{k_1}{k_2}}\frac{2 k_2}{3 k_1+3 k_2}
\label{eqsme}
\end{eqnarray}
\begin{figure}
\centering{\psfig{figure=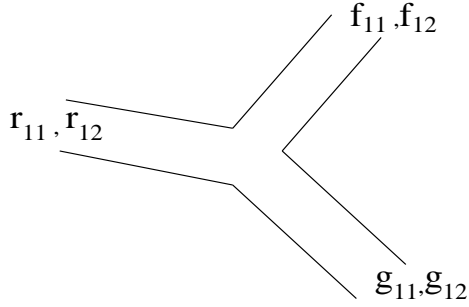,width=6cm,height=4cm,angle=0}}
\caption{A $3$-leg junction.}\label{figleg}
\end{figure}
Mode matching at the junction $P$ [Fig. \ref{figsys}] gives
\begin{eqnarray}
\left ( \begin{array}{c}
r^\prime_{11} \\ r^\prime_{12} \\ g^\prime_{11} \\ g^\prime_{12} \\ A \\ C
\end{array} \right ) &=& S_J \left ( \begin{array}{c}
1 \\ 0 \\ 0 \\ 0 \\ B \\ D
\end{array} \right )\label{eqp}
\end{eqnarray}
Similarly, mode matching at the junction $Q$ [Fig. \ref{figsys}] gives
\begin{eqnarray}
\left ( \begin{array}{c}
Be^{-ik_1l_1} \\ De^{-ik_2l_1} \\ E \\ G \\ Ke^{-ik_1l_3} \\ Me^{-ik_2l_3}
\end{array} \right ) &=& S_J \left ( \begin{array}{c}
Ae^{ik_1l_1} \\ Ce^{ik_2l_1} \\ Fe^{-i\alpha} \\ He^{-i\alpha} \\ Je^{i(k_1l_3+\beta)} \\ Le^{i(k_2l_3+\beta)}
\end{array} \right )\label{eqq}
\end{eqnarray}
\noindent
Here $l_1$, $l_2$ and $l_3$ are shown in Fig. \ref{figsys}.
$\alpha+\beta=2\pi\phi/\phi_0$, $\phi$ being the magnetic flux and
$\phi_0=hc/e$ is the flux quantum. Eqn. \ref{eqq} automatically applies
periodic boundary conditions to wave function in the ring. Mode matching at the
impurity site $X$ [Fig. \ref{figsys}] gives
\begin{eqnarray}
\left( \begin{array}{c} F e^{-ik_1l_2}\\H e^{-ik_2l_2}\\J\\L \end{array}\right) & = & 
\left( \begin{array}{cccc}
\tilde r_{11} & \tilde r_{12} & \tilde t_{11} & \tilde t_{12}\\\tilde r_{21} & 
\tilde r_{22} & \tilde t_{21} & \tilde t_{22}\\
\tilde t_{11} & \tilde t_{12} & \tilde r_{11} & \tilde r_{12}\\\tilde t_{21} &
 \tilde t_{22} & \tilde r_{21} & \tilde r_{22}\\
\end{array}\right) \nonumber\\
& & \times \left( \begin{array}{c} E e^{i(k_1l_2+\alpha)}\\ G e^{i(k_2l_2+\alpha)} \\K e^{-i\beta} \\M e^{-i\beta} \end{array}\right)\label{eqx}
\end{eqnarray}
where ~\cite{pfb90},
\begin{eqnarray}
\tilde r_{pp^\prime} &=& \frac{-i\frac{\Gamma_{pp^\prime}}{2\sqrt{k_pk_{p^\prime}}}}{1+\sum_e \frac{\Gamma_{ee}}{2\kappa_e}+ i\sum_p\frac{\Gamma_{pp}}{2k_p}}
\end{eqnarray}
$\sum_e$ represents sum over all the evanescent modes and $\sum_p$ represents
sum over all the propagating modes. $p$ or $p^\prime$ can take values 1 and 2
as there are two propagating modes. 
$\kappa_e=\sqrt{\frac{e^2\pi^2}{W^2}-\frac{2mE}{\hbar^2}}$,
where $e=3,4,..$. The inter-mode (i.e. $p\ne p^\prime$) transmission amplitudes
are  $\tilde t_{pp^\prime}=\tilde r_{pp^\prime}$ and intra-mode transmission
amplitudes are $\tilde t_{pp}=1+\tilde r_{pp}$. $\Gamma_{pp^\prime}$ is given
as
\begin{eqnarray}
\Gamma_{pp^\prime} &=& \frac{2m\gamma}{\hbar^2} \sin [\frac{p\pi(q+W/2)}{W}] \sin [\frac{p^\prime\pi(q+W/2)}{W}]\nonumber\\
\end{eqnarray}
where $q$ is used to denote the position coordinate of the $\delta$-potential
impurity.
\vskip .2in
We calculate $A$, $B$, $C$, $D$, $E$, $F$, $G$, $H$, $J$, $K$, $L$ and $M$
numerically from Eqn. \ref{eqp}, Eqn. \ref{eqq} and Eqn. \ref{eqx}.
Persistent current is defined by
\begin{eqnarray}
I &=& \int_{-\frac{W}{2}}^{\frac{W}{2}} \frac{\hbar}{2im}(\Psi^\dag\vec{\bigtriangledown}\Psi-\Psi\vec{\bigtriangledown}\Psi^\dag) dy
\label{eqiwf}
\end{eqnarray}
which can be simplified to give
\begin{eqnarray}
I &=& I^{(k_1)}+I^{(k_2)}
\label{eqik1k2}
\end{eqnarray}
\noindent
Here
\begin{eqnarray}
I^{(k_1)} &=& 2 I_0 (|E|^2-|F|^2+|G|^2-|H|^2)^{(k_1)}
\label{eqik1}
\end{eqnarray}
\noindent
is the current when electron is incident along the left lead in $k_1$
channel. This is the scattering problem defined by Eqn. \ref{eqp}, Eqn.
\ref{eqq} and Eqn. \ref{eqx}. Similarly,
\begin{eqnarray}
I^{(k_2)} &=& 2 I_0 (|E|^2-|F|^2+|G|^2-|H|^2)^{(k_2)}
\label{eqik2}
\end{eqnarray}
\noindent
is the current when electron is incident along the left lead in $k_2$
channel and this scattering problem has to be solved by using a similar
set of equations. Here, $I_0=\frac{\hbar e}{2mW^2}$.
\begin{figure}
\centering{\psfig{figure=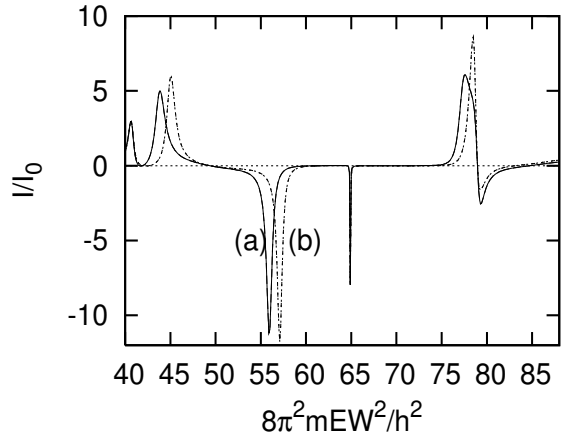,width=8cm,height=6cm,angle=0}}
\caption{$I/I_0$ vs $8\pi^2mEW^2/h^2$ with (a) $\gamma=0$ (solid line) and (b)
$\gamma=4$ (dashed line). The system parameters are $l_1=l_2=l_3=1$,
$\alpha=\beta=0.3$.} \label{figprc}
\end{figure}
\vskip .2in
The nature of current obtained from Eqn. \ref{eqik1k2} is shown in
Fig \ref{figprc}. We take the energy range
($4 \pi^2 \le \frac{2mEW^2}{\hbar^2} \le 9 \pi^2$, i.e.
$40 \le \frac{2mEW^2}{\hbar^2} \le 88$) in such a way that both the modes are
propagating. This gives the behavior that is captured in earlier works
\cite{gm90}$^,$\cite{va90}$^,$\cite{as91}. The single ring current can be
paramagnetic as well as diamagnetic as can be seen from Fig. \ref{figprc}.
We shall show that when we make one of the modes evanescent, we will get a
behavior that is not discussed before. As soon as $2mEW^2/\hbar^2$ becomes less
than $4\pi^2$, the second channel becomes evanescent. This is because $k_1$ is
real, whereas $k_2$ is imaginary ($k_2\rightarrow i\kappa_2$ in this regime).
No electron can be incident into an evanescent channel from infinity and so
$I^{(k_2)}$ will not exist. But electrons incident in $k_1$ channel can be
excited into an evanescent channel in the ring implying that $G$ and $H$ in
Eqn. \ref{eqik1} are non-zero. A single impurity can excite an electron into
the evanescent second channel. Scattering at the junctions can also excite an
electron into the evanescent second channel. An electron residing in an
evanescent state will carry a current. This naturally arises in the scattering
problem that is defined in Eqns. \ref{eqp}-\ref{eqx}. Evanescent mode current
can be calculated by directly applying Eqn. \ref{eqiwf} to evanescent mode
wave-functions or it can be calculated by analytically continuing propagating
mode current to below the barrier. Both results are equal.
\vskip .2in 
The $S$ matrix becomes $2\times 2$ and is given by
\begin{eqnarray}
S=\left ( \begin{array}{cc}
r^\prime_{11} & g^\prime_{11}\\ g^\prime_{11} & r^\prime_{11}
\end {array} \right )
\end{eqnarray}
\begin{figure}
\centering{\psfig{figure=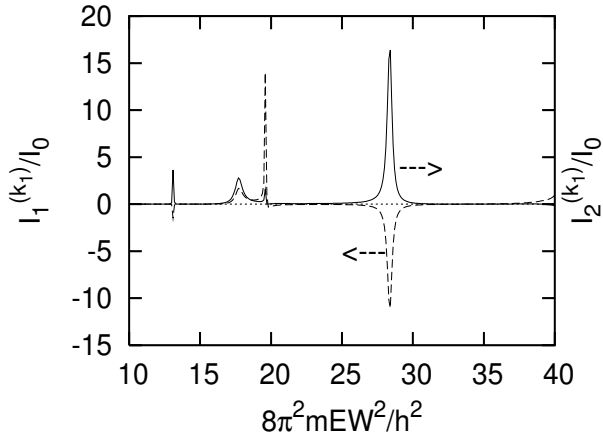,width=8cm,height=6cm,angle=0}}
\caption{$I^{(k_1)}_1/I_0$ and $I^{(k_1)}_2/I_0$ vs $8\pi^2mEW^2/h^2$. The
system parameters are $l_1=l_2=l_3=1$, $\alpha=\beta=0.3$, $\gamma=4$.}
\label{figi1i2}
\end{figure}
Although the $S$ matrix is $2 \times 2$, its calculation has to be done by
using the $6 \times 6$ junction matrix $S_J$ defined in Eqn. \ref{eqsmj} and
the $4 \times 4$ impurity $S$-matrix defined in Eqn. \ref{eqx}.
$g^\prime_{12}$, $r^\prime_{12}$ etc are still non-zero, although they do not
carry any current but they define the coupling to the evanescent mode. So Eqns.
\ref{eqp}, \ref{eqq} and \ref{eqx} still holds with $k_2\rightarrow i\kappa_2$
where $\kappa_2=\sqrt{\frac{4\pi^2}{W^2}-\frac{2mE}{\hbar^2}}$ . Unitarity
should imply $\mid r^\prime_{11} \mid^2+\mid g^\prime_{11} \mid^2=1$ and indeed
we get this from the junction matrix defined by $S_J$ in Eqn. \ref{eqsmj} and
impurity $S$-matrix defined in Eqn.\ref{eqx}. This implies that $S_J$ is
appropriate to account for realistic multichannel situations. $S_U$ does not
take into account such effects and does not allow one to include coupling to
evanescent modes, maintaining unitarity. Current is expected to be continuous
as the energy changes continuously from evanescent modes to propagating modes
at $\frac{2mEW^2}{\hbar^2}=4\pi^2$. This also comes out in our calculations
\begin{figure}
\centering{\psfig{figure=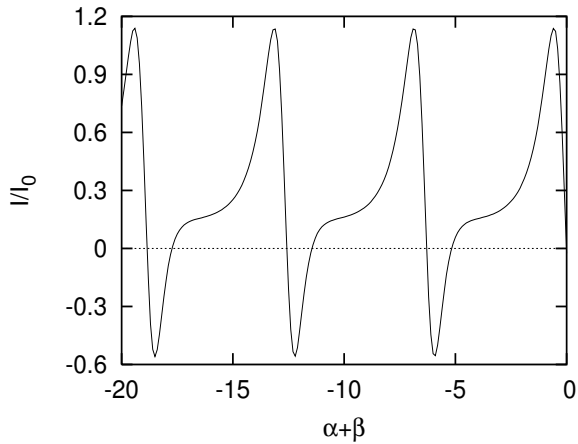,width=8cm,height=6cm,angle=0}}
\caption{$I/I_0$ vs $\alpha$+$\beta$ when second channel is evanescent.
Here $l_1=1.0$, $l_2=0.2$, $l_3=1.8$ and $\gamma=-4.11$.}
\label{figif}
\end{figure}
\vskip .2in
Note from Eqn. \ref{eqik1} that $I^{(k_1)}=I^{(k_1)}_1+I^{(k_1)}_2$, where
$I^{(k_1)}_1=2 I_0 (\mid E \mid^2-\mid F\mid^2)^{(k_1)}$, $E$ and $F$ being the
wave function amplitudes in the propagating channel and
$I^{(k_1)}_2=2 I_0 (\mid G \mid^2-\mid H\mid^2)^{(k_1)}$, $G$ and $H$ being the
wave-function amplitudes in the evanescent channel. In Fig. \ref{figi1i2} we
have plotted $I^{(k_1)}_1$ and $I^{(k_1)}_2$ versus $2mEW^2/\hbar^2$. While
$I^{(k_1)}_1$ can be positive (diamagnetic) as well as negative (paramagnetic),
$I^{(k_1)}_2$ is seen to be only diamagnetic. $I^{(k_1)}_2$ is the current in
an evanescent channel, and there is a fundamental difference with evanescent
channel currents in 1D. In 1D evanescent channel current cannot oscillate with
energy because evanescent wave-function is not of wave nature \cite{psd96}. In
1D we have to introduce an infinitesimal region of the ring where the electron
can be propagating (evanescent in the rest of the ring), for the persistent
current to be oscillating between paramagnetism and diamagnetism \cite{cb03}.
But in the present case the second channel is evanescent throughout the length
of the ring, its wave-function is not of wave nature, and yet can oscillate
with Fermi energy. The peaks in $I^{(k_1)}_1$ are resonance effects due to wave
nature of electron wave-function wherein at these energies, the electrons can
spend a long time in the propagating mode. The impurity also gets a long time
to pump more electrons into the evanescent mode. So the evanescent mode current
$I^{(k_1)}_2$ also peaks at the same energies where $I^{(k_1)}_1$ peaks (see
Fig. \ref{figi1i2}) although the evanescent mode wave-function is not of wave
nature. The difference between them being that while the peaks in $I^{(k_1)}_1$
can be in positive direction (diamagnetic) or in negative direction
(paramagnetic), the peaks in $I^{(k_1)}_2$ are, always in the positive
direction. As impurity configuration changes, these peaks also change randomly.
But the peaks in $I^{(k_1)}_2$ always follow the peaks in $I^{(k_1)}_1$. One
can also see this mathematically. Although the evanescent mode wave-function is
not of a wave nature, $G$ and $H$ are functions of $k_1$ and $\kappa_2$, due to
the non-locality of quantum mechanics. $E$ and $F$ are also functions of $k_1$
and $\kappa_2$. While $I^{(k_1)}_1$ will fluctuate around zero value,
$I^{(k_1)}_2$ will fluctuate around a certain positive value as disorder
configuration changes. Apart from this shift, $I^{(k_1)}_2$ will follow the
same rules as $I^{(k_1)}_1$ as far as disorder averaging is concerned. Or more
appropriately, $I^{(k_1)}_2$ will follow same averaging rules as conductance
that fluctuate with disorder, remaining positive all the time. It is much
easier to take random values of $l_2$ and $l_3$ to show this for the average
current.
\vskip .2in
The observable current $I=I^{(k_1)}$ when second channel is evanescent, is
plotted versus flux in Fig. \ref{figif}. The figure shows that when a
diamagnetic component is present, the response looks like that observed in
experiments done by Deblock et al \cite{rdrb02}. One can further check the
validity of our explanation by measuring how the response of the ensemble
scales with the number of rings in the ensemble. One has to go to a large
enough ensemble so that the first harmonic has averaged to a flux independent
diamagnetic component. This component will scale linearly with $N$, the number
of rings present in the ensemble, while the flux dependent part will scale as
$\sqrt{N}$.
\vskip .2in
Solution of Schr\"{o}dinger equation in multichannel rings consist of
evanescent modes that are naturally populated due to scattering. These
evanescent modes can carry large persistent current that are diamagnetic in
nature and are as sensitive to disorder as propagating modes. Previous attempts
to explain the experimental results on persistent current ignore their
contribution as they were thought to be small and insensitive to disorder. They
provide a natural explanation for the discrepency between theory and
experiments. Future experiments should try to isolate the role of the
contributions coming from propagating and evanescent modes.

The authors would like to acknowledge useful discussions with Prof.
A.M. Jayannavar. One of us (PSD) would like to acknowledge useful
discussions with Prof. H. Bouchiat.

\end{document}